\documentclass[aps,amssymb,twocolumn,superscriptaddress,unsortedaddress,nofootinbib]{revtex4}
\usepackage{setspace}
\usepackage{amsfonts}
\usepackage{amsmath}
\usepackage{multirow}
\usepackage{graphicx}
\usepackage{psfrag}
\usepackage[dvips]{color}

\def\be{\begin{equation}}
\def\ee{\end{equation}  }
\def\bea{\begin{eqnarray}}
\def\eea{\end{eqnarray}  }

\begin{document}
\title{Black hole formation in Randall-Sundrum II braneworlds}

\author{Daoyan Wang}
\affiliation{6282 Kathleen Avenue, Suite 204, 
             Burnaby, British Columbia, 
             V5H 4J4 Canada}
\author{Matthew W. Choptuik}
\affiliation{
     Department of Physics and Astronomy,
     University of British Columbia,
     Vancouver, British Columbia, V6T 1Z1 Canada}
\affiliation{CIFAR Cosmology and Gravity Program, 
	     180 Dundas St W, Suite 1400, 
             Toronto, Ontario, M5G 1Z8 Canada}

\begin{abstract}
We present the first numerical study of the full dynamics of a  braneworld scenario,
working within the framework of the single brane model of Randall and Sundrum (RSII).
In particular, we study the process of gravitational collapse driven by a massless 
scalar field which is confined to the brane. 
Imposing spherical symmetry on the brane,  
we show that the evolutions of  
sufficiently strong initial configurations of the scalar field result in black holes
that have finite extension into the bulk. Furthermore, we find preliminary evidence that 
the black holes generated form a unique sequence, irrespective of the details of 
the initial data. 
The black hole solutions we obtain from dynamical evolutions are 
consistent with those previously computed from a static vacuum ansatz.

\end{abstract}

\maketitle

\noindent \textbf{\textit{I. Introduction.}} 
The basic idea of braneworld scenarios is that our observable universe could be 
a 3+1 dimensional brane embedded in a higher dimensional bulk spacetime.
The second  such model invented by Randall and Sundrum, known as RSII, 
is a 4+1-dimensional braneworld containing a single brane on which all  matter is confined,
and one extra dimension of infinite size into which only gravity can propagate~\cite{RSIII}. 
This model is remarkable for its simplicity and 
recovers general relativity (GR) on the brane in the weak field regime~\cite{BraneEinstein}
even though the extra dimension is of infinite extent.
However, the behaviour of the model in the context of strong, dynamical gravitational fields is 
not as clear and is the focus of this letter.

One vacuum solution of the RSII model is~\cite{RSIII, CHawking99} 
\begin{equation}
{\rm d}s^2=\frac{\ell^2}{z^2}\left(h_{ab}{\rm d}x^a {\rm d}x^b +{\rm d}z^2\right)~~\text{where~}z\ge \ell. \label{bg} 
\end{equation}
Here, $z$ labels the extra dimension, with the 
single brane of the RSII model located at $z=\ell$,
$h_{ab}$ is the metric of a Ricci-flat solution
to Einstein's equations in 4 dimensions~\cite{CHawking99}, and
$x^a$, $a=0,1,2,3$ are the coordinates on the brane.
When $h_{ab}$ is the Minkowski metric, 
the spacetime described by (\ref{bg}) is 
a part of a Poincar\'{e} patch of an anti-de Sitter (AdS) spacetime with 
AdS length $\ell$~\cite{HawkingBook}. 
When $h_{ab}$ is a 4-dimensional black hole metric, (\ref{bg}) 
describes a black string~\cite{CHawking99}. 
However, due to the Gregory-Laflamme instability~\cite{GLunstable},
we do not expect that black strings will form from generic gravitational collapse
of matter on the brane and a natural question that arises is: what are the end states
of such collapse processes?

Noting that the Gregory-Laflamme instability is most severe at the AdS horizon 
where black strings might ``pinch off'',
Chamblin, Hawking and Reall  
proposed that collapse would yield a black object with finite extension 
into the bulk, i.e.~a black hole~\cite{CHawking99}.
Many groups have since investigated one aspect of the Chamblin-Hawking-Reall
proposal---the existence of such black holes---using both analytical and
numerical approaches (see~\cite{largeBH} and references therein).
One issue that became a key topic of debate in these studies was whether 
or not black holes with sizes
large relative to the AdS scale, $\ell$, could even exist.
Eventually, Figueras and Wiseman~\cite{largeBH} obtained 
black holes with sizes in the approximate range $[0.07\ell, \,20\ell]$ 
(then $\sim$ $[5\times10^{-4} \ell ,\, 100\ell]$ 
in~\cite{largeBH3}) by numerically constructing the solutions 
of static vacuum RSII spacetimes.  Their technique involved 
perturbing ${\rm AdS_5/CFT_4}$ solutions, which themselves were
numerically constructed~\cite{AdS5CFT4}. 
Additionally, black holes, including ones with large sizes,  were obtained numerically by 
Abdolrahimi et al~\cite{largeBH2}. 
Although they used a different computational approach than Figueras and Wiseman, they also assumed 
that the spacetimes were static and vacuum, and the two sets of results are presumably consistent. 

Without proving the stability and uniqueness of these solutions, 
however, it is not clear whether they describe the end states of 
matter collapse on the brane. 
A direct way to address this and other issues 
is to solve numerically the full set of dynamical Einstein-matter equations for a simple collapse 
process, and this is what is described in the remainder of this paper.
To our knowledge, this is the first numerical work 
that treats the full dynamics of a braneworld scenario with physical branes. 
Due to the prohibitive computational cost of performing the 
calculations in the fully 5-dimensional context, we start with one of the simplest 
possible setups in which a black hole could form. This involves
(a) imposing spherical symmetry on the brane, which makes the 
bulk axisymmetric, and (b) using a massless scalar field
as the matter source confined on the brane. 

\noindent \textbf{\textit{ II. Methodology.}} 
For convenience, we define the ``background'' spacetime as
that described by~(\ref{bg}) with $h_{ab}$ being the Minkowski metric.
We study gravitational collapse in the RSII spacetimes that asymptotically
go to this background at all spatial infinities. 
It is known that Cauchy surfaces exist in the RSII spacetimes~\cite{CHawking99}, 
and the evolution can thus be formulated as an initial value problem
with Einstein's equations in the bulk providing the governing equations 
for the gravitational field.
On the brane the scalar field satisfies its usual 3+1 equation of motion.  
Additionally, the brane and the matter confined to it are coupled to 
the bulk through Israel's junction conditions~\cite{ref:Israel}.
We use coordinates 
$(t,r,\theta,\phi,z)$ adapted to the axisymmetry in the bulk: 
once spherical symmetry on the brane is imposed the coordinate system is 
effectively cylindrical, with dynamical variables
depending on $(t,r,z)$, and we thus refer to it as such.
The numerical calculations are carried out using finite difference approximation and employ
coordinates $\hat R=r/(r+r_0)$ and $\hat Z=(z-\ell)/(z-\ell+z_0)$, 
where $r_0$ and $z_0$ are adjustable parameters, which compactify the spatial domain.
We adopt the generalized harmonic formalism of GR~\cite{GHcartoon2} which yields a 
strongly hyperbolic set of evolution equations and, correspondingly,
a well-posed initial value problem. 

To utilize existing numerical relativity techniques regarding 
coordinate choices, which have typically been applied only to
asymptotically flat spacetimes, we perform a conformal transformation 
so that the conformally transformed spacetimes are asymptotically flat. 
Denoting the metric of the physical spacetime as $g_{\mu\nu}$, 
the conformally transformed metric is defined as 
$\tilde g_{\mu\nu} \equiv \Psi^{-2}g_{\mu\nu}$, 
where the conformal function $\Psi$ goes to $\ell/z$ at spatial infinity. 
Under the conformal transformation the field equations have the form of 
Einstein's equations in terms of $\tilde g_{\mu\nu}$,
with an additional term related to $\Psi$~\cite{lecture} that
can be treated as a matter contribution, from the perspective of $\tilde g_{\mu\nu}$.

Due to our use of cylindrical coordinates we must exercise care to minimize irregularities
in the numerical solutions at and near the axis of symmetry.  
Our approach 
is to first carry the Cartesian components
of the various tensors and pseudo-tensors appearing in our scheme into cylindrical coordinates
via coordinate transformation relations.  We then let these components
serve as fundamental dynamical variables in the numerical calculations. 
The interested reader is directed to~\cite{thethesis} for a complete 
description of this methodology, as well as many other additional 
details of the calculations.
In terms of such  
components the most generic form of the metric in axisymmetry can be written
\begin{equation}
		\tilde g_{\alpha\beta} = \left(
\begin{array}{ccccc}
  \tilde\eta_{tt} & \tilde\eta_{tr} & 0 & 0 & \tilde\eta_{tz}\\
  \tilde\eta_{tr} & \tilde\eta_{rr} & 0 & 0 & \tilde\eta_{rz}\\
  0 & 0 &\tilde\eta_{\theta\theta}r^2 & 0 & 0 \\
  0 & 0 & 0 &\tilde\eta_{\theta\theta}r^2\sin^2\theta& 0 \\
  \tilde\eta_{tz} & \tilde\eta_{rz} & 0 & 0 &\tilde\eta_{zz}
\end{array}
\right). \label{g}
\end{equation}
The boundary conditions at the symmetry axis ($r=0$) are 
$\tilde \eta_{tt,r}=\tilde \eta_{rr,r}=\tilde \eta_{zz,r}=\tilde \eta_{tz,r}=\tilde \eta_{tr}=\tilde \eta_{rz}=0$. 
The metric function
$\tilde \eta_{\theta\theta}$ is rewritten as
$\tilde \eta_{rr}+r \tilde W$, so that the local flatness
condition~\cite{localFlat, thethesis} 
$\tilde \eta_{\theta\theta}|_{r=0}=\tilde \eta_{rr}|_{r=0}$ is automatically
satisfied. The parity condition $\tilde \eta_{\theta\theta,r}\big|_{r=0}=0$ 
then becomes $W|_{r=0}=0$. 
Similarly to~(\ref{g}), the source functions of the generalized harmonic formalism,
$\tilde H_\mu = -\tilde g_{\mu\nu}
\tilde \Gamma^\nu_{\,\ \alpha\beta}\tilde g^{\alpha\beta}$, are expressed in terms
of Cartesian components, $\tilde h_\mu$, as
\begin{equation}
\left(
\begin{array}{ccccc}
  \tilde H_t \\
  \tilde H_r \\
  \tilde H_\theta \\
  \tilde H_\phi\\
  \tilde H_z
\end{array}
\right) = \left(
\begin{array}{cccc}
 \tilde h_t+(2/r)\left(\tilde\eta_{tr}/\tilde\eta_{\theta\theta}\right) \\
 \tilde h_r+(2/r)\left(\tilde\eta_{rr}/\tilde\eta_{\theta\theta}\right) \\
 \cot \theta \\
  0 \\
 \tilde h_z+(2/r)\left(\tilde\eta_{rz}/\tilde\eta_{\theta\theta}\right)
\end{array}
\right).\label{h}
\end{equation}
The coordinate conditions are now imposed via the $\tilde h_\mu$.  
The $\tilde\eta_{\mu\nu}$, $\tilde W$ and $\tilde h_\mu$ constitute the full set of
fundamental variables in our simulations.

The brane imposes interesting new physics, as well as new challenges 
for numerical calculations.
During the evolutions, while the damping term introduced in~\cite{GHdamping} suffices
to damp the constraint-violating modes in the bulk,
it can not control the corresponding modes appearing at and near the brane.
We solve this problem by explicitly enforcing the constraints at the brane.
These constraints can be converted into conditions on 
$\left.\tilde \eta_{\mu z,z}\right|_{z=\ell}$, 
and can therefore be treated as boundary conditions.
Generically there are no boundary conditions 
for $\tilde \eta_{\mu z}$~\cite{thethesis} at the brane, but 
by utilizing the coordinate freedom there
we impose the additional boundary conditions 
$\left.\tilde\eta_{tz}\right|_{z=\ell}=\left.\tilde\eta_{rz}\right|_{z=\ell}=0$,
so that apparent horizons and the brane are perpendicular
where they intersect~\cite{billbob}. 

\begin{figure}
\begin{center}
\includegraphics[height=3.6in,draft=false]{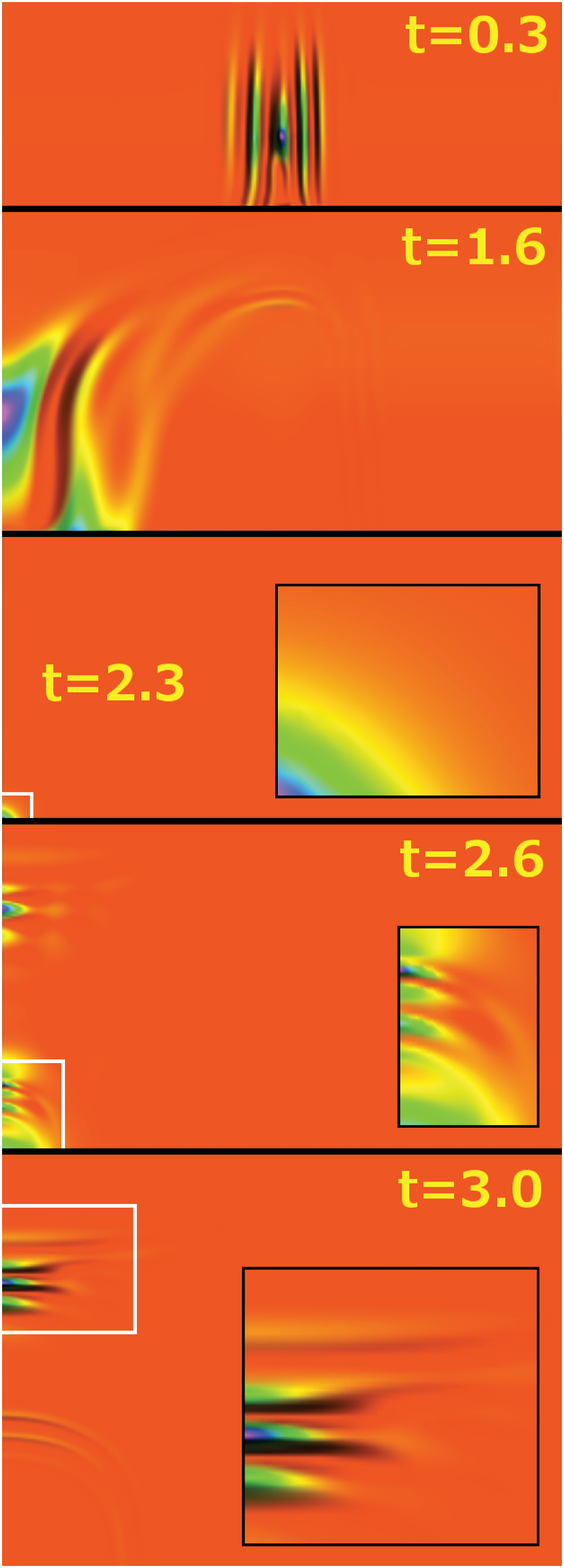}
\includegraphics[height=3.6in,draft=false]{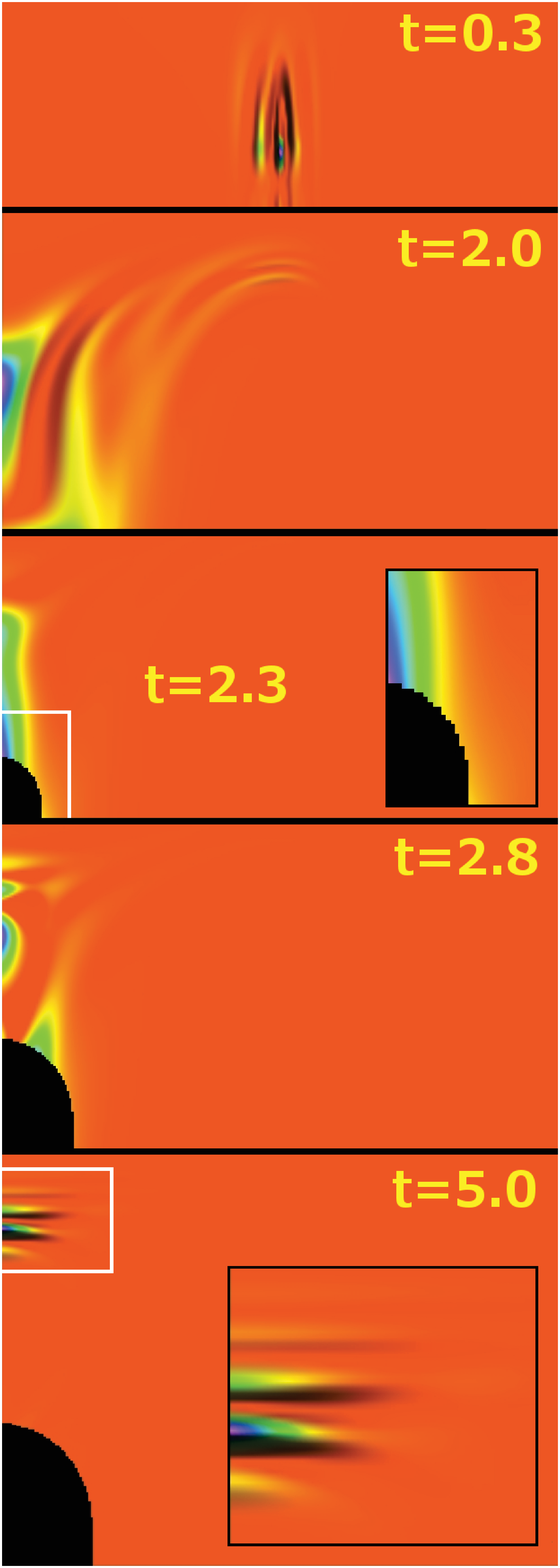}
\end{center}
\caption{Time evolution of the Kretschmann scalar,
$R_{\mu\nu\alpha\beta}R^{\mu\nu\alpha\beta}$, from 
two calculations with initial data having  parameters 
$(A_0,x_0,\sigma_r)=(0.04,2,0.2)$ (left panel) and $(0.24,2,0.2)$ (right panel), 
respectively.
Evolution proceeds top to bottom. The time label is coordinate time, $t$,
and the horizontal and vertical axes are the compactified computational 
coordinates $\hat R$ and $\hat Z$ with compactification 
parameters $(r_0,z_0)=(2,2)$, so that the bottom of each panel is $z=\ell$.  
The evolution on the left results in complete 
dispersal of the scalar field, while that on the right generates a black hole 
with $r_{\rm a}=0.61\ell$. The black ellipses in the bottom three frames of 
the right sequence are regions interior to apparent horizons. 
The insets (outlined in black) are magnified views of portions of the corresponding frames 
(outlined in white) and 
are designed to highlight various features of the calculations.
}
\label{fig:evo}
\end{figure}

\noindent \textbf{\textit{ III. Results.}} 
Our scheme can be used to study a wide range of dynamical processes
but here we focus attention on the end states of gravitational collapse.
Our expectation is that these end states will be stationary and, as will
be seen, there is fairly strong evidence that this is the case.
Our initial data for the massless scalar field on the brane, $z=\ell$,
is a localized Gaussian pulse, given by
$\Phi(0,r)=A_0 \cdot \exp\left[-(r-x_0)^2/\sigma_r^2\right]$, 
where $A_0, x_0$ and $\sigma_r$ are adjustable parameters.  
We further specify the initial data to be time symmetric, 
so the pulse evolves into distinct ingoing and outgoing pieces. 
For weak data, and completely analogously to the 4 dimensional GR case, the ingoing pulse 
implodes through $r = 0$,
then propagates outwards to infinity (Fig.\ref{fig:evo}, left panel).
For strong enough initial data, on the other hand, the ingoing pulse 
becomes sufficiently self-gravitating that an apparent horizon with finite 
extension into the bulk forms (Fig.\ref{fig:evo}, right panel).
We use this as a signal that the resulting spacetime contains a black hole, 
also with finite extension into the bulk.
Once an apparent horizon is detected in a calculation, we implement 
black hole excision~\cite{unruh}, which enables us to continue the evolution
for many dynamical times.

\begin{figure}
\begin{center}
\includegraphics[height=1.9in,draft=false]{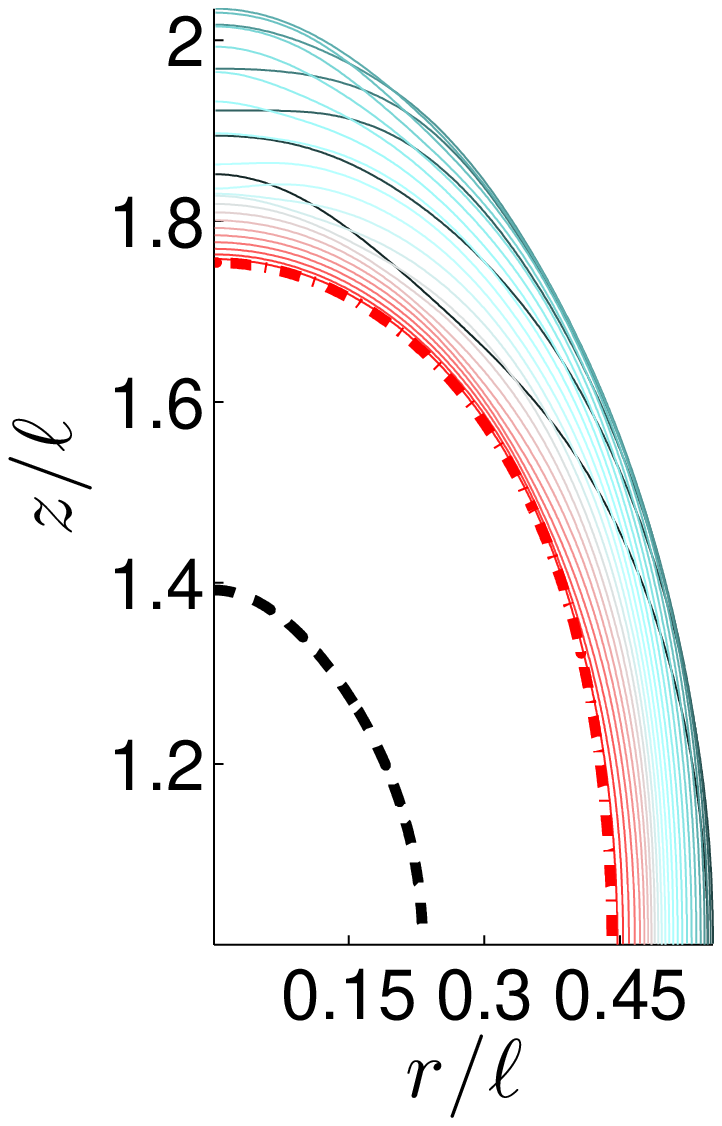}~~
\includegraphics[height=1.9in,draft=false]{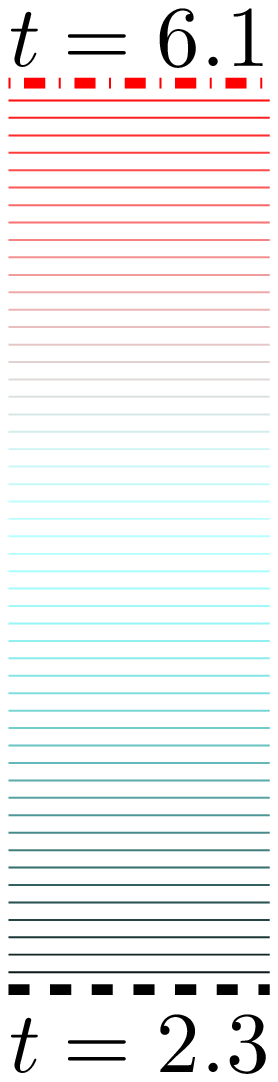}~~
\includegraphics[height=1.9in,draft=false]{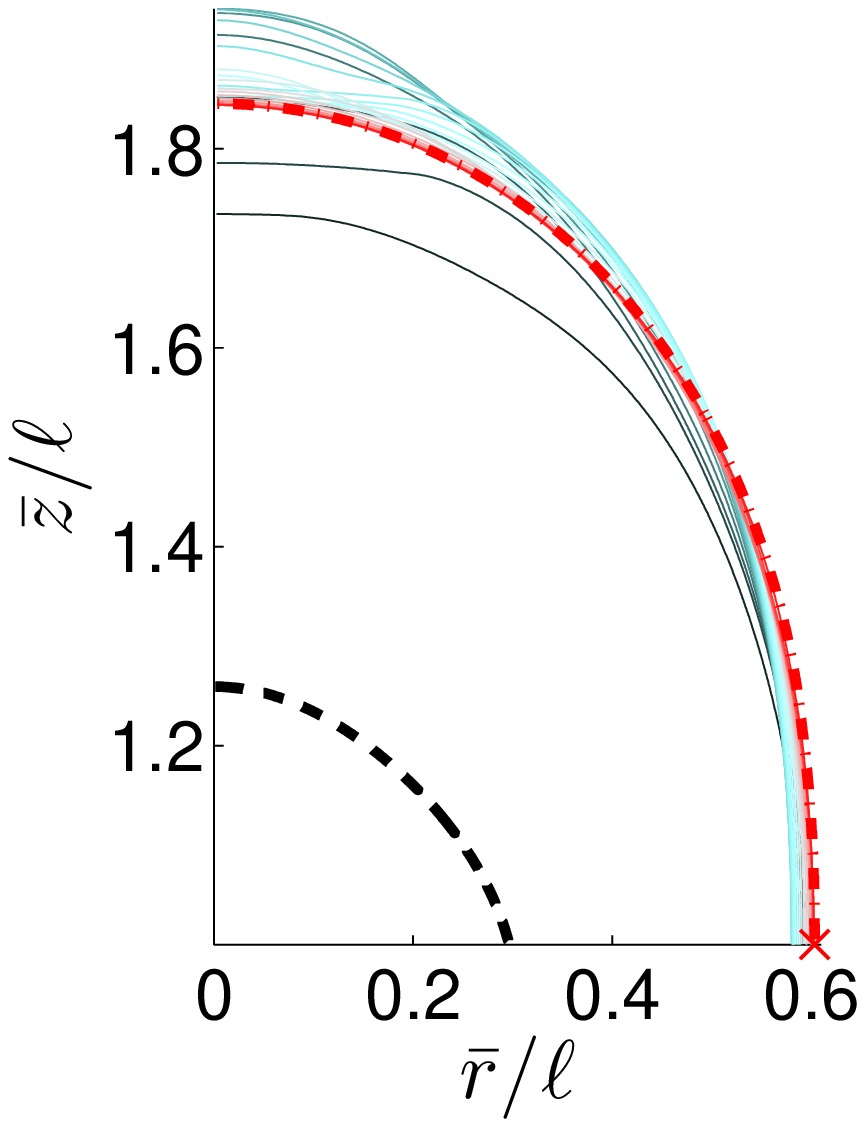}
\end{center}
\caption{Evolution of the apparent horizon for initial conditions 
$(A_0,x_0,\sigma_r)=(0.24,2,0.2)$.  
The left panel shows the evolution in coordinate space $(r,z)$ while the right panel 
shows the time development of the apparent horizons in the background space whose coordinates
are denoted $(\bar r,\bar z)$, and where the embedding of the horizons in the background preserves
their intrinsic geometries. The middle panel defines the 
legend for the line colors and line types. 
The black dashed line denotes the first apparent horizon 
which appeared at $t=2.3$, while the red dotted-dashed line is the last horizon computed at $t=6.1$. 
Intermediate horizons are represented by lines with colors changing continuously
from black to red as the evolution progresses. 
The embedding graph (right panel) 
shows that the apparent horizons converge to a 
limiting line, which represents the physical process
of the black hole settling into an apparently stationary state.
The intersection of the brane with the horizon corresponding to this state
is marked with an $\times$ in the graph (lower right).
From the definition of $r_{\rm a}$ the coordinates of the intersection
are ($\bar r,\bar z)=(r_{\rm a},\ell)$.
}
\label{fig:evo2}
\end{figure}

Our main analysis of the features of the black holes that form 
is made through the structure of the apparent horizons in the bulk. 
As is well known, the apparent horizon does not generally coincide with the event horizon.
However, if an apparent horizon exists in a stationary, regular predictable spacetime, 
it does agree with the event horizon~\cite{HawkingBook}. 
Furthermore, when there is no matter in the vicinity of a horizon, 
the intersection of the bulk event horizon with the brane can be shown to be a well-defined 
event horizon on the brane~\cite{EHPRL}. 
Since our spacetimes are axisymmetric, the intersection of an 
apparent horizon with the brane is spherically symmetric on the brane and the intersection's 
proper area, $\mathcal A_{\rm brane}$, defines an areal radius,
$r_{\rm a}\equiv \sqrt{\mathcal A_{\rm brane}/4\pi}$, 
which is called the {\it size} of the black hole~\cite{largeBH}.  
During the calculations we monitor $r_{\rm a}$ as well as  
the proper area of the horizon in the bulk, $\mathcal A_{\rm bulk}$, 
and the proper circumference of the horizon extended into the bulk, $C_5$.
To further analyze the apparent horizons we generate embedding diagrams in the 
background space that preserve the intrinsic geometries of the horizons~\cite{thethesis}.  
Fig.~\ref{fig:evo2} shows sample time developments of the apparent horizon displayed 
both in coordinate space and using embedding diagrams, for the same calculation 
visualized in the right panel of Fig.~\ref{fig:evo}.

We are interested in the long term behaviour of the black hole solutions and, 
in particular, the extent to which stationary end states are achieved.
If the slicing condition is such that the $t={\rm const}$ slices are 
Lie-dragged by the Killing vector associated with the putatively stationary 
end state, the intrinsic geometry of the apparent horizon
will itself be time-independent, resulting in a stationary embedding plot.
Unfortunately, such a slicing condition is non-trivial to devise and we have yet 
to formulate one.  However, during a typical long term evolution of 
a collapsing configuration, where the elapsed integration time is much greater than
the characteristic dynamical time scale, we find that the solution settles 
into some {\em apparently stationary state} that we use to approximate 
the true end state. 
We note that the evolution inevitably departs from this configuration.
We assume that this is a coordinate effect but have not yet been able
to develop a technique to show that this is the case.
Operationally we identify the apparently stationary states 
by looking for approximate time-independence in the embedding diagrams, as 
illustrated in the right panel of Fig.~\ref{fig:evo2}.

\begin{figure}
\begin{center}
\includegraphics[height=2.28in,draft=false]{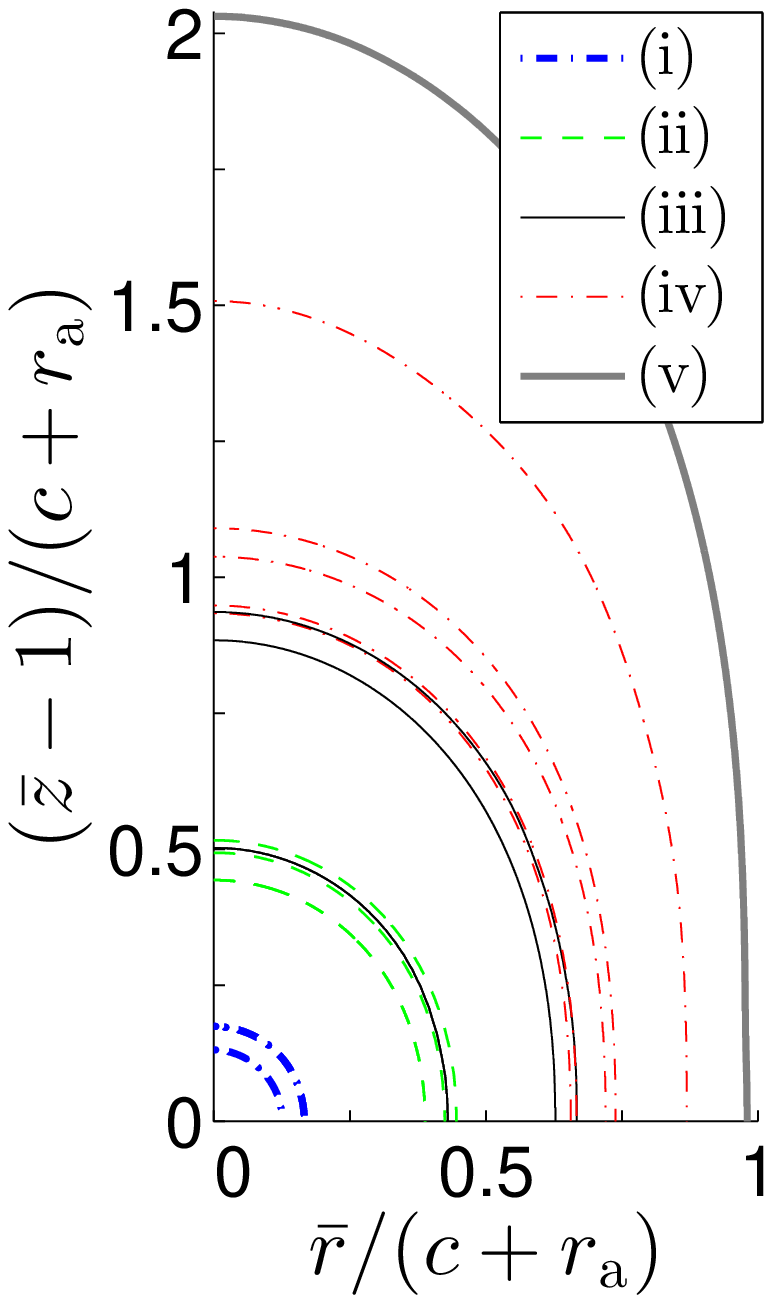}~~
\includegraphics[height=2.48in,draft=false]{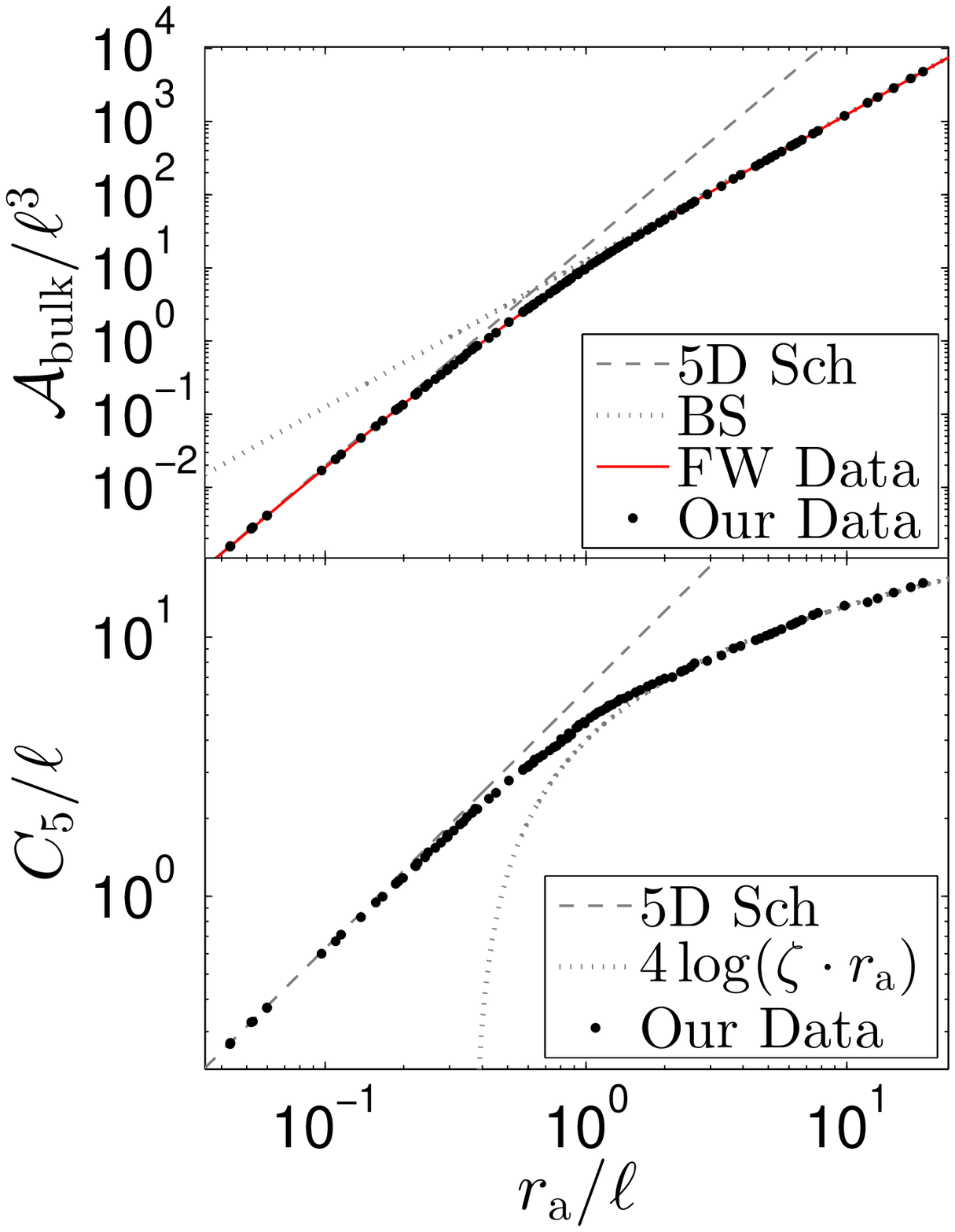}
\end{center}
\caption{
Left: Apparently stationary black holes 
generated from different initial data families, visualized using 
the embedding of apparent horizons in the 
background space whose coordinates are $(\bar r, \bar z)$,
and where the compactification parameter is $c=0.3\ell$.
This plot provides evidence that the black holes generated from distinct 
initial data families form a single, unique sequence. Additionally, 
the sequence appears to be ``monotonic''
in the sense that for initial data strengths $A_0$ and $A'_0$, where $A_0' > A_0$,
the horizon embedding curve for $A'_0$ lies entirely outside of that for $A_0$.
Three black holes from family (iii) are shown here, 
the smallest fits precisely between the two largest of family (ii), while 
the largest is well fit by the smallest of family (iv).
Upper right: $\mathcal A_{\rm bulk}$ as a function of $r_{\rm a}$, with
a comparison with the data obtained by 
Figueras and Wiseman (FW)~\cite{largeBH}. 
Small black holes asymptote to their 5-dimensional
Schwarzschild (5D Sch) counterparts and 
the $\mathcal A_{\rm bulk}$-vs-$r_{\rm a}$ relation of 
large black holes asymptotes to that of black strings (BS).
Lower right: $C_5$ as a function of $r_{\rm a}$.  
}
\label{fig:nohair}\label{fig:fw}\label{fig:C5}
\end{figure}

We have performed a series of evolutions using five distinct initial data families, 
from which we have obtained black holes with  sizes in the range
$0.04\ell \lesssim r_{\rm a} \lesssim 19.6\ell$. Within each family 
only $A_0$ is varied; $x_0$ and $\sigma_r$
are held fixed. The specific values of $(x_0,\sigma_r)$ for the families are: 
(i)~$(0.5,0.1)$; (ii)~$(1,0.1)$; (iii)~$(2,0.2)$; (iv)~$(0,0.3)$;  (v)~$(2,0.5)$. 
The apparently stationary states for the complete set of evolutions 
are plotted in the left panel of Fig.~\ref{fig:nohair},
where the axes have been scaled in order to accommodate the significant range of sizes 
of the black holes.
The figure shows preliminary evidence that 
the specific black holes in the RSII spacetimes studied here---which are
axisymmetric in the bulk, and which have no angular 
momentum or non-gravitational charges---comprise a unique sequence,
regardless of the details of the initial data.
Furthermore, since the sequence is monotonic in the sense defined in the figure caption,
we can use $r_{\rm a}$ to define the position of any given black hole in the sequence.
If the black hole solutions are unique, the black holes we obtain 
from dynamical evolution should agree with those previously obtained
via a static vacuum ansatz by Figueras and Wiseman~\cite{largeBH, largeBH3}.
The comparison with their results is shown in the upper right panel of Fig.~\ref{fig:fw} 
which plots $\mathcal A_{\rm bulk}$ as a function of $r_{\rm a}$.  
Clearly the agreement is very good.

When the sizes of the black holes are much smaller than $\ell$, the solutions tend to
5-dimensional asymptotically flat Schwarzschild ones~\cite{largeBH}.
As shown in the left panel of Fig.~\ref{fig:nohair}, as the size of the black hole increases, 
its embedding-diagram shape changes from spherical to prolate (cigar-shaped).
Now, large black holes have an 
upper limit which is the corresponding AdS$_5$/CFT$_4$ solution~\cite{largeBH, thethesis}.
The geometry of the portion of a large black hole that is close to 
the brane behaves like that of a black string, which makes 
the $\mathcal A_{\rm bulk}$-versus-$r_{\rm a}$ relation 
gradually change into that of black strings.  
To study the difference between large black holes and black strings,
we plot $C_5$-versus-$r_{\rm a}$
in the lower right panel of Fig.~\ref{fig:C5}.
Again, we see that small black holes are asymptotically
5-dimensional Schwarzschild, so $C_5 \sim 2\pi r_{\rm a}$. For large black holes, 
$C_5$ is well approximated by 
$C_5\sim4\log(\zeta \cdot r_{\rm a})$~\cite{FWp},
as long as the {\it shape} of the holes (as seen in an embedding plot),
does not change with the size~\cite{thethesis}. 
For our data we find a best fit $\zeta\approx 2.71$ while
Figueras and Wiseman~\cite{FWp} independently and previously obtained 
$\zeta\approx 2.8$  from their static vacuum calculations.  
The proper circumference $C_5$ increases with $r_{\rm a}$, but 
the ratio $C_5/r_{\rm a}$ shrinks as $\sim\log r_{\rm a}/ r_{\rm a}$. Therefore
large black holes are actually prolate 
(flattened pancakes) as determined from their intrinsic geometry.
This was first suggested in~\cite{BulkShape}.

\noindent{\bf{\em IV. Conclusion and Discussion.}}
We have performed the first numerical study of the full dynamics
of a braneworld within the framework of the RSII model. 
We find that the result of gravitational collapse of a strong 
pulse of massless scalar field is a black hole with finite extension into the bulk.
There is preliminary evidence that the black holes that form constitute
a unique sequence that can be conveniently parameterized by the areal 
radius of the horizon.  Additionally, the black hole solutions that we compute 
have properties in agreement with those found previously from a static vacuum
ansatz~\cite{largeBH, largeBH3}.

Our approach could be improved through the 
development of a slicing condition that would adapt to 
the assumed time translational symmetry of the stationary end states. 
Black holes with size $0.04\ell \lesssim r_{\rm a} \lesssim 19.6\ell$
were constructed here. 
Consideration of a wider range of sizes would also 
require improvement of coordinate conditions. 
It would be particularly useful to be able to probe the 
regime of small black holes since the nature of 
black hole critical phenomena in this scenario could well be different
than it is in usual 4-dimensional spacetime. 

\noindent{\bf{\em  Acknowledgments.}}
This research was supported by NSERC and CIFAR. 
We thank Evgeny Sorkin and William Unruh for useful discussions, and 
Toby Wiseman and Pau Figueras for providing their data and discussing
it with us.
Some of the calculations were performed using the Westgrid facilities 
of Compute Canada.

\end{document}